\documentclass[ALICE,manyauthors]{cernphprep}

\usepackage[comma,square,numbers,sort&compress]{natbib}
\usepackage{hyperref}
\usepackage{lineno}

\usepackage{multirow}
\usepackage{tabularx}
\usepackage{graphicx}
\usepackage{amsmath}
\usepackage{mathptmx}      
\usepackage{bm}
\usepackage{subfigure}
\usepackage{amssymb}
\usepackage{amsmath}

\newcommand{\pp}{p\mbox{--}p}
\newcommand{\pL}{p\mbox{--}$\mathrm{\Lambda}$}
\newcommand{\pXim}{p\mbox{--}$\mathrm{\Xi^-}$}
\newcommand{\LL}{$\mathrm{\Lambda}$\mbox{--}$\mathrm{\Lambda}$}

\newcommand{\ppColl}{\ensuremath{\mathrm {p\kern-0.05em p}}}
\newcommand{\pPbColl}{\ensuremath{\mathrm{p\mbox{--}Pb}}}
\newcommand{\EbinResult}{$B_{\Lambda\Lambda} = 3.2^{+1.6}_{-2.4}\text{(stat)}^{+1.8}_{-1.0}\text{(syst)}$~MeV}
\newcommand{\EbinResultSTAT}{$B_{\Lambda\Lambda} = 3.2^{+1.6}_{-2.4}\text{(stat)}$~MeV}

\begin{document}%

\begin{titlepage}
\PHyear{2019}
\PHnumber{096}      
\PHdate{10 May}  
%

\title{Study of the $\mathrm\mathbf{\Lambda}$\mbox{--}$\mathrm\mathbf{\Lambda}$ interaction  with femtoscopy correlations in $\mathrm{\mathbf {p\kern-0.05em p}}$ and $\mathrm{\mathbf{p\mbox{--}Pb}}$ collisions at the LHC}
\ShortTitle{Detailed study of the {\LL} interaction  with femtoscopy in small systems}   

\Collaboration{ALICE Collaboration\thanks{See Appendix~\ref{app:collab} for the list of collaboration members}}
\ShortAuthor{ALICE Collaboration} 

\begin{abstract}
This work presents new constraints on the existence and the binding energy of a possible {\LL} bound state, the H-dibaryon, derived from {\LL} femtoscopic measurements by the ALICE collaboration. The results are obtained from a new measurement using the femtoscopy technique in {\ppColl} collisions at $\sqrt{s}=13$~TeV and {\pPbColl} collisions at $\sqrt{s_{\mathrm{NN}}}=5.02$~TeV, combined with previously published results from {\ppColl} collisions at $\sqrt{s}=7$~TeV.
The {\LL} scattering parameter space, spanned by the inverse scattering length $f_0^{-1}$ and the effective range $d_0$, is constrained by comparing the measured {\LL} correlation function with calculations obtained within the Lednick\'y model. The data are compatible with hypernuclei results and lattice computations, both predicting a shallow attractive interaction, and permit to test different theoretical approaches describing the {\LL} interaction.
The region in the $(f_0^{-1},d_0)$ plane which would accommodate a {\LL} bound state is substantially restricted compared to previous studies. The binding energy of the possible {\LL} bound state is estimated within an effective-range expansion approach and is found to be \EbinResult.
\end{abstract}
\end{titlepage}
\setcounter{page}{2}

\section{Introduction and physics motivation}
\noindent
A detailed characterization of the {\LL} interaction is of fundamental interest since it plays a decisive role in the quantitative understanding of the hyperon (Y) appearance in dense neutron-rich matter, in proto-neutron and in neutron stars \cite{Oertel:2016bki}. 
If hyperons do appear at large densities and their fraction becomes sizeable, the Y\mbox{--}Y interaction is expected to play an important role in the equation of state of the system \cite{SchaffnerBielich:2000yj,Weissenborn:2011ut}.
Even if the hyperon densities in compact objects are negligible, the interplay between the average separations and the {\LL} effective range determine the possible onset of phenomena such as fermion superfluidity, and hence influence the transport properties of the system \cite{Carlson:2012mh,Guven:2018sgo,Tanigawa:2002ex}.

The characterization of the {\LL} interaction is still an open issue in experimental nuclear physics. The Nagara event, recently measured with the emulsion technique \cite{Takahashi:2001nm,Ahn:2013poa}, reports a clear evidence for a double-$\Lambda$ hypernucleus $\mathrm{_{\Lambda\Lambda}^{6} He}$, with a small binding energy between the two $\Lambda$s of $\Delta B_{\Lambda\Lambda}=\, 0.67 \pm 0.17$ MeV.
This value was obtained by comparing the binding energy of the two $\Lambda$s inside the double hypernucleus ($B_{\Lambda\Lambda}=\, 6.91 \pm 0.16$ MeV) with the binding energy of a single $\Lambda$ in a single-hypernucleus, however, it might be influenced by three-body forces.
Nevertheless, this result was used to set a lower limit for the mass of the predicted but so far not observed H-dibaryon, a possible bound state composed of six quarks (uuddss) \cite{Jaffe:1976yi}. Several experimental collaborations have been involved in the search for this state in the decay channels $\mathrm{H\rightarrow \Lambda p \pi}$ and $\mathrm{H\rightarrow \Lambda \Lambda}$, in nuclear and elementary ($\mathrm{e^-e^+}$) collisions, but no evidence has been found \cite{Adam:2015nca,Kim:2013vym,Chrien:1998yt}, even though an enhanced {\LL} production near threshold was measured by E224 and E522 at KEK-PS \cite{Yoon:2007aq,Ahn:1998fj}.

Theoretical models constrained to the available nucleon--nucleon and hyperon-nucleon experimental data, assuming either a soft~\cite{Maessen:1989sx,Rijken:1998yy,Ueda:1998bz} or a hard~\cite{Nagels:1976xq,Nagels:1978sc} repulsive core for the {\LL} interaction, predict different scattering lengths ($f_0$) and effective ranges ($d_0$). Throughout this paper the standard sign convention in femtoscopy is used, according to which a positive $f_0$ corresponds to an attractive interaction, while a negative scattering length corresponds either to a repulsive potential ($d_0>|f_0|/2$) or a bound state ($d_0<|f_0|/2$). It was reported that a small variation of the {\LL} repulsive core parametrization leads to inverse scattering lengths within $-0.27\,\,\mathrm{fm}^{-1} < f_0^{-1} < 4\,\,\mathrm{fm}^{-1} $ and effective ranges up to 16 fm~\cite{Morita:2014kza}. Other calculations are directly constrained to the Nagara event and result in rather small scattering lengths and moderate effective ranges, like the FG $(f_0^{-1}=\,1.3 \,\,\mathrm{fm}^{-1};\, d_0 =\,6.59 \,\,\mathrm{fm})$ \cite{Filikhin:2002wm} and the HKMYY $(f_0^{-1}= \,1.74 \,\,\mathrm{fm}^{-1};\,d_0= \, 6.45 \,\,\mathrm{fm})$ \cite{Hiyama:2002yj} models. It is clear that more experimental data are needed to study the problem in a more quantitative and model-independent way.

An alternative method to study hypernuclei is the investigation of momentum correlations of {\LL} pairs produced in hadron--hadron collisions via the femtoscopy technique \cite{Lisa:2005dd}. The STAR collaboration reported a {\LL} scattering length and effective range of $f_0^{-1}=\,-0.91\pm 0.31^{+0.07} _{-0.56} \,\, \mathrm{fm}^{-1}$ and $d_0=\,8.52\pm 2.56^{+2.09} _{-0.74} \,\, \mathrm{fm}$, measured in Au\mbox{--}Au collisions at $\sqrt{s_\mathrm{NN}}=200$ GeV \cite{Adamczyk:2014vca}. These values correspond to a repulsive interaction; however, it was shown that the values and the sign of the scattering parameters strongly depend on the treatment of feed-down contributions from weak decays to the measured correlation. A re-analysis of the data outside the STAR collaboration came to different conclusions \cite{Morita:2014kza} and resulted in a shallow attractive interaction.

In a pioneering study \cite{RUN1}, the {\LL} interaction was studied employing the femtoscopy technique in {\ppColl} collisions at $\sqrt{s}=7$ TeV. This study demonstrated that the data are consistent with either a bound state or an attractive interaction, however, due to the small data sample no quantitative results were obtained. In this letter, these studies are extended by analyzing final-state momentum correlations in {\ppColl} collisions at $\sqrt{s}=13$ TeV and {\pPbColl} collisions at $\sqrt{s_\mathrm{NN}}=5.02$~TeV, recorded by ALICE during LHC Run 2. The small system size in {\ppColl} and {\pPbColl} gives rise to pronounced correlations from strong final-state interactions due to the small relative distance at which particles are produced. Hence, the large data sets enable a high-precision study of the {\LL} strong final-state interaction and provide new experimental constraints on the scattering parameters and the existence of a possible bound state. 
\section{Data analysis}\label{sec:Exp}
\noindent
The analysis presented in this paper is based on the data samples collected by ALICE~\cite{alice} during the Run 2 of the LHC (2015--2018) in {\ppColl} collisions at $\sqrt{s}=$13~TeV and {\pPbColl} collisions at $\sqrt{s_\mathrm{NN}}=5.02$~TeV, combined with the previously analyzed Run 1 data from {\ppColl} collisions at $\sqrt{s}=$7~TeV~\cite{RUN1}. The event and particle candidate selection criteria follow closely the procedure applied in the Run 1 analysis~\cite{RUN1}.

The events are triggered using two V0 detectors, which are small-angle plastic scintillator arrays placed on either side of the collision vertex at pseudorapidities $2.8 < \eta < 5.1$ and $-3.7 < \eta < -1.7$ \cite{Abelev:2014ffa}.
Minimum bias {\ppColl} and {\pPbColl} events are triggered by the requirement of coincident signals in both V0 detectors, synchronous with the beam crossing time defined by the LHC clock.
The V0 detector is also used to reject background events stemming from the interaction of beam particles with the beam pipe materials or beam-gas interactions. Pile-up events with more than one collision per bunch crossing are rejected by evaluating the presence of secondary event vertices \cite{Abelev:2014ffa}. Charged particles are reconstructed by the Inner Tracking System (ITS)~\cite{alice} and the Time Projection Chamber (TPC)~\cite{2010TPCNIMA}, both immersed in a 0.5~T solenoidal magnetic field directed along the beam axis. A uniform detector coverage is assured by requiring the maximal deviation between the reconstructed primary vertex (PV) and the nominal interaction point to be smaller than 10~cm. 
The PV can be reconstructed with the combined information of the ITS and TPC, and independently with the Silicon Pixel Detector (SPD - one of the three subdetectors of the ITS). If both methods are available, the difference of the $z$-coordinate between both vertices is required to be smaller than 5\,mm.
After applying these selection criteria the remaining number of events is $1.0 \times 10^{9}$ for the {\ppColl} at $\sqrt{s}=$13~TeV sample and $6.1 \times 10^{8}$ for {\pPbColl} at $\sqrt{s_\mathrm{NN}}=5.02$~TeV. This corresponds to about 90\,\% and 84\,\% of all processed events in  {\ppColl} and {\pPbColl}.

The {\LL} interaction is the main focus of the present study. As will be explained in the next section, the {\pp} correlation function is an essential input for the femtoscopic analysis of {\LL}. Therefore, the reconstruction of both protons and $\Lambda$ particles will be described in the following paragraphs. To increase the statistical significance of the result, the anti-particle pairs are measured as well.

The selection of the proton candidates follows the analysis strategy used for the {\ppColl} collisions at $\sqrt{s}=$7~TeV~\cite{RUN1}. The particle identification (PID) is determined by the number of standard deviations $n\sigma$ between the hypothesis for a proton and the experimental measurement of the specific energy loss $\mathrm{d}E/\mathrm{d}x$ in the TPC or the timing information from the Time-Of-Flight (TOF) detector \cite{Akindinov2013}. The analyzed tracks are selected within the kinematic range $0.5<p_\mathrm{T}<4.05$~GeV/$c$ and $|\eta|<0.8$. The PID is performed only with the TPC for tracks with $p<0.75~$GeV/$c$, by requiring $|n\sigma|<3$. To maintain the purity of tracks with $p>0.75~$GeV/$c$, the $|n\sigma|$ is calculated from combining the TPC and TOF information. The contribution of secondary particles, which stem from electromagnetic and weak decays or the detector material, are a contamination in the signal. The fractions of primary and secondary protons are extracted using Monte Carlo (MC) template fits to the distance of closest approach of the particles to the PV~\cite{Aamodt:2010dx}. The MC distributions are generated using Pythia 8.2 \cite{PYTHIA} for the {\ppColl} and DPMJET 3.0.5 \cite{Roesler:2000he} for the {\pPbColl} case, filtered through the ALICE detector and reconstruction algorithm \cite{alice}. The proton purity in {\ppColl} ({\pPbColl}) is found to be 99 (97)\% with a primary fraction of 85 (86)\%.

The $\mathrm{\Lambda}$ particles are reconstructed via the decay 
$\mathrm{\Lambda\rightarrow p\pi^-}$, which has a branching ratio of 63.9\% and $c\tau=7.89~$cm~\cite{PDG}. For the reconstruction of the $\overline{\Lambda}$ the charge conjugate decay is employed.
The interaction rate of the LHC varied during different periods of the {\ppColl} running. 
To maintain a constant purity that is independent of the interaction rate, in addition to the selection criteria used for the analysis of {\ppColl} collisions at $\sqrt{s}=$7~TeV~\cite{RUN1}, the charged decay tracks must either have a hit in one of the SPD or Silicon Strip Detector (SSD - ITS subdetector) layers or a matched TOF signal.
After applying all selection criteria the final $\Lambda$ and $\overline{\Lambda}$ candidates are selected in a $4$~MeV/$c^{2}$ ($\sim 3 \sigma$) mass window around the nominal mass~\cite{PDG}. 
The fractions of primary and secondary $\Lambda$ particles are extracted similarly as the protons, while the observable for the template fits is the cosine of the opening angle $\alpha$ between the $\Lambda$  momentum and the vector pointing from the PV to the $\Lambda$ decay vertex. The $\Lambda$ purity in {\ppColl} ({\pPbColl}) is found to be 97 (94)\% with a primary fraction of 59 (50)\%. The exact composition of secondaries, as well as the  $\Lambda$ to $\Sigma^0$ ratio, is fixed in the MC simulations, but is model dependent. Therefore, the systematic uncertainties include a $20\%$ variation of the ratios of these contributions.
\section{Analysis of the correlation function}\label{sec:Ana}

The method used to investigate the {\LL} interaction relies on particle pair correlations measured as a function of $\vec{k^*}$, defined as the single-particle momentum in the pair rest frame~\cite{Lisa:2005dd}. The observable of interest $C(\vec{p_1},\vec{p_2})$ is defined as the ratio of the probability of measuring simultaneously two particles with momenta $\vec{p_1}$ and $\vec{p_2}$, to the product of the single-particle probabilities:
\begin{equation}\label{eq:CkStat}
C(\vec{p_1},\vec{p_2})=\frac{P(\vec{p_1},\vec{p_2})}{P(\vec{p_1})P(\vec{p_2})}.
\end{equation}
In the absence of correlations, the numerator factorizes and the correlation function becomes unity. The femtoscopy formalism~\cite{Lisa:2005dd} relates the correlation function for a pair of particles, to their effective two-particle emitting source function $S(r)$ and the two-particle wave function $\Psi(\vec{k^*},\vec{r})$:
\begin{equation}\label{eq:Ck}
C(k^*)=\int S(r) \mid \Psi(\vec{k^*},\vec{r})\mid ^2\mathrm{d}^3 r\xrightarrow{k^*\rightarrow\infty}1,
\end{equation}
where $r$ is the relative distance between the points of emission of the two particles. This definition of $C(k^*)$ assumes that the emission source is not dependent on $k^*$, it is spherically symmetric and the emission of all particles is simultaneous. 
The EPOS transport model \cite{Pierog:2013ria} predicts an emission source that does not fully satisfy the above assumptions. However, it was verified that the above simplifications result in very mild deviations in the correlation functions, which are negligible for the present analysis.

For a spherical symmetric potential the angular dependence of the wave-function is trivially integrated out. Thus the direction of $\vec{k^*}$ becomes irrelevant on the left-hand side of Eq.~\ref{eq:Ck}. Particles with large relative momenta $q^*=2k^*$ are not correlated, leading to $C(k^*\rightarrow\infty)=1$. 

The strong interaction has a typical range of a few femtometers and thus a significant modification of the wave function with respect to its asymptotic form is expected only for $r\lesssim2~$fm. Consequently, for small emission sources the correlation function will be particularly sensitive to the strong interaction potential. Experimentally, a small emission source can be formed in {\ppColl} and {\pPbColl} collisions \cite{RUN1,CATS}. In the current analysis, it is assumed that the emission profile is Gaussian and that the {\pp} and {\LL} systems are characterized by a common source size $r_0=r_\text{\pp}=r_\text{\LL}$, which is determined by fitting the {\pp} correlation function and then used for the investigation of the {\LL} interaction. In {\ppColl} collisions the effect of mini-jets is only present for baryon correlations between particle and anti-particle \cite{RUN1}, hence the investigated data provide a clean environment to extract the femtoscopic signal.

Two different frameworks are available for the computation of $C(k^*)$. The first tool used in this analysis is the ``Correlation Analysis Tool using the Schr\"odinger equation'' (CATS) \cite{CATS}. Here, a local potential $V(r)$ is used as the input to a numerical evaluation of the wave function and the corresponding correlation function. CATS delivers an exact solution and this tool is used to model the {\pp} correlation using a Coulomb and an Argonne $v_{18}$ potential \cite{Wiringa:1994wb} for the strong interaction. The known {\pp} interaction allows the source size $r_0$ to be extracted from the fit to the measured correlation function.

The second tool is the Lednick\'y model \cite{Lednicky:1981su}, which assumes a Gaussian emission source and evaluates the wave function in the effective-range expansion. In this approach, the interaction is parameterized in terms of the scattering length $f_0$ and the effective range $d_0$. This approach produces a very accurate approximation for $C(k^*)$ in case $d_0\lesssim r_0$, while for smaller values of $r_0$ the approximate solution may become unstable, in particular for negative values of $f_0$ \cite{RUN1}. However, it is known that the Lednick\'y model can be used to model the {\pL} correlation function even for a source size of $r_0=1.2~$fm, with a deviation from the exact solution of less than $4\%$ \cite{CATS}. It is therefore expected that this model can successfully be used to study the {\LL} interaction, even in small collision systems. Nevertheless, the validity of the approximation will be further verified in the next section.

Experimentally, the correlation function is defined as
\begin{equation}\label{eq:ExpCk}
C_{\text{exp}}(k^*)=\mathcal{N}\frac{N_{\text{same}}(k^*)}{N_{\text{mixed}}(k^*)}\xrightarrow{k^*\rightarrow\infty}1,
\end{equation}
where $N_{\text{same}}(k^*)$ and $N_{\text{mixed}}(k^*)$ are the same and mixed event distributions, while
$\mathcal{N}$ is a normalization constant determined by the condition that particle pairs with large relative momenta are not correlated. In small collision systems $C_{\text{exp}}(k^*)$ often has a long-range tail due to momentum conservation, and a related approximately linear non-femtoscopic background extending to low $k^*$ \cite{RUN1}. The latter is incorporated by including a linear term in the fit function.

To increase the statistical significance of $C_{\text{exp}}(k^*)$ the particle-particle (PP) and antiparticle-antiparticle ($\mathrm{\overline{P}\overline{P}}$) correlations are combined using their weighted mean $C_{\text{exp}}(k^*) = \mathcal{N}_\text{PP}C_{\text{exp,PP}}(k^*) \oplus \mathcal{N}_\mathrm{\overline{P}\overline{P}}C_\mathrm{exp,\overline{P}\overline{P}}(k^*)$, with the normalization performed in the range $240<k^*<340$~MeV/$c$, which is unaffected by femtoscopic correlations. It was verified that $\mathcal{N}_\text{PP}C_{\text{exp,PP}}(k^*) = \mathcal{N}_\mathrm{\overline{P}\overline{P}}C_\mathrm{exp,\overline{P}\overline{P}}(k^*)$ within the statistical uncertainties.

The systematic uncertainties of the experimental correlation function are evaluated by varying the selection criteria of the proton and $\Lambda$ candidates within 20\%, following the procedure used for the analysis of the {\ppColl} collisions at $\sqrt{s}=$7~TeV~\cite{RUN1}. Nevertheless, by performing a Barlow test~\cite{Barlow}, the systematic uncertainties were found to be insignificant compared to the statistical uncertainties.

Momentum resolution effects modify the correlation function by at most $10\%$ and are accounted for by correcting the theoretical correlation function \cite{RUN1}. The measured experimental correlation function contains not only the correlation signal of interest, but additionally accumulates residual contributions from feed-down particles. These are considered in the theoretical description of the correlation by using the linear decomposition of the total correlation function into
\begin{equation}\label{eq:decomp}
C_\mathrm{tot}(k^*)=\sum_i \lambda_i C_i(k^*),    
\end{equation}
where the sum runs over all contributions, the $\lambda$ parameters are the weight factors for the different contributions to the total correlation and $i=0$ corresponds to the primary correlation. The $\lambda$ coefficients are determined in a data-driven approach by performing Monte Carlo template fits to the data, using Pythia and DPMJET in {\ppColl} and {\pPbColl} collisions, respectively. The values obtained are summarized in Table~\ref{tab:lambdaval}. The systematic uncertainties are determined from the variation of the composition of secondary contributions, and the $\Lambda$ to $\Sigma^0$ ratio.
\begin{table*}
\begin{center}
\begin{tabularx}{\textwidth}{l|X|X || l|X|X || l|X|X || l|X|X}
\hline  \hline
\multicolumn{3}{c||}{\pp} &
\multicolumn{3}{c||}{\pL} &
\multicolumn{3}{c||}{\pXim} &
\multicolumn{3}{c}{\LL} \\

\multirow{2}{*}{Pair}  & $\lambda_i^{\ppColl}$  & $\lambda_i^{\pPbColl}$ & 
\multirow{2}{*}{Pair}  & $\lambda_i^{\ppColl}$  & $\lambda_i^{\pPbColl}$ & 
\multirow{2}{*}{Pair}  & $\lambda_i^{\ppColl}$  & $\lambda_i^{\pPbColl}$ & 
\multirow{2}{*}{Pair} & $\lambda_i^{\ppColl}$ & $\lambda_i^{\pPbColl}$ \\  
&  (\%) &  (\%) &  
&  (\%) &  (\%) &
&  (\%) &  (\%) &
&  (\%) &  (\%) \\
\hline 
pp & 74.8 & 72.8 &
p$\Lambda$ & 50.3 & 41.5 &
p$\Xi^-$ & 55.5 & 50.8 &
$\Lambda \Lambda$ & 33.8 & 23.9\\
pp$_\Lambda$ & 15.1 & 16.1 &
p$\Lambda_{\Sigma^0}$ & 16.8 & 13.8 &
p$\Xi^-_{\Xi(1530)^-}$ & 8.8 & 8.1 &
 & &\\
 & & &
p$\Lambda_{\Xi^-}$ & 8.3 & 12.1 &
 & &
 & &\\
flat res. & 8.1 & 8.0 &
flat res. & 20.4 & 24.9 &
flat res. & 30.3 & 28.3 &
flat res. & 59.8 & 64.0\\
fakes & 2.0 & 3.1 &
fakes & 4.2 & 7.7 &
fakes & 5.4 & 12.8 &
fakes & 6.4 & 12.1 \\
\hline \hline
\end{tabularx}
\caption[lambda parameters]{The weight parameters (Eq.~\ref{eq:decomp}) $\lambda_i^{\ppColl}$ and $\lambda_i^{\pPbColl}$ of the individual components of the {\pp}, {\pL}, {\pXim} and {\LL} correlation functions. The sub-indexes are used to indicate the mother particle in case of feed-down. Only the non-flat feed-down (residual) contributions are listed individually, while all other  contributions are listed as ``flat residuals (res.)''. All misidentified (fake) pairs are assumed to be uncorrelated, thus resulting in a flat correlation signal.}
\label{tab:lambdaval}
\end{center}
\end{table*}
The individual contributions $C_i(k^*)$ are modeled either using CATS or the Lednick\'y model. The non-genuine ($i\ne0$) contributions include additional kinematic effects which lead to a smearing of their corresponding correlation functions~\cite{Kisiel:2014mma}. As the correlation strength of these residuals is strongly damped one can assume that $C_{i\ne 0}(k^*)\approx 1$~\cite{OliThesis}. The only significant contribution is {\pL}$\rightarrow${\pp}, where the {\pL} interaction is modeled using the scattering parameters from a next-to-leading order (NLO) $\chi$EFT calculation~\cite{Haidenbauer:2013oca} and the corresponding correlation function is computed using the Lednick\'{y} model. The remaining residuals are considered flat, apart from  {\pXim}$\rightarrow${\pL}, $\mathrm{p}$\mbox{--}$\Sigma^0\rightarrow${\pL} and $\mathrm{p}$\mbox{--}$\Xi(1530)^-\rightarrow${\pXim}, where the interaction can be modeled. For the {\pXim} interaction a recent lattice QCD potential, from the HAL QCD collaboration~\cite{Hatsuda:2017uxk, Sasaki:2017ysy}, is used. The $\mathrm{p}$\mbox{--}$\Sigma^0$ is modeled as in~\cite{Stavinskiy:2007wb}, while $\mathrm{p}$\mbox{--}$\Xi(1530)^-$ is evaluated by taking only the Coulomb interaction into account.

\begin{figure*}[h]
\centering{
\includegraphics[width=0.49\textwidth]{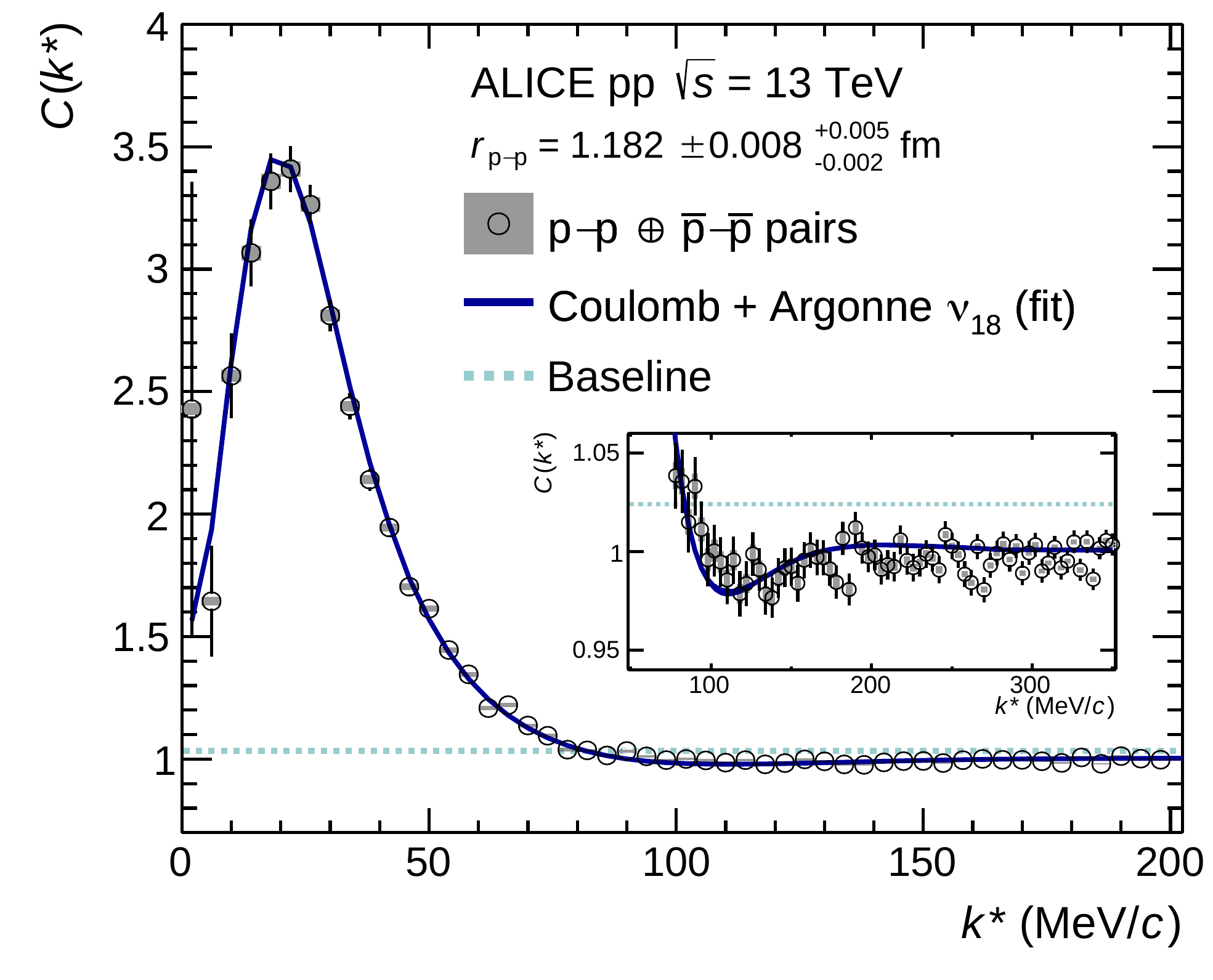}
\includegraphics[width=0.49\textwidth]{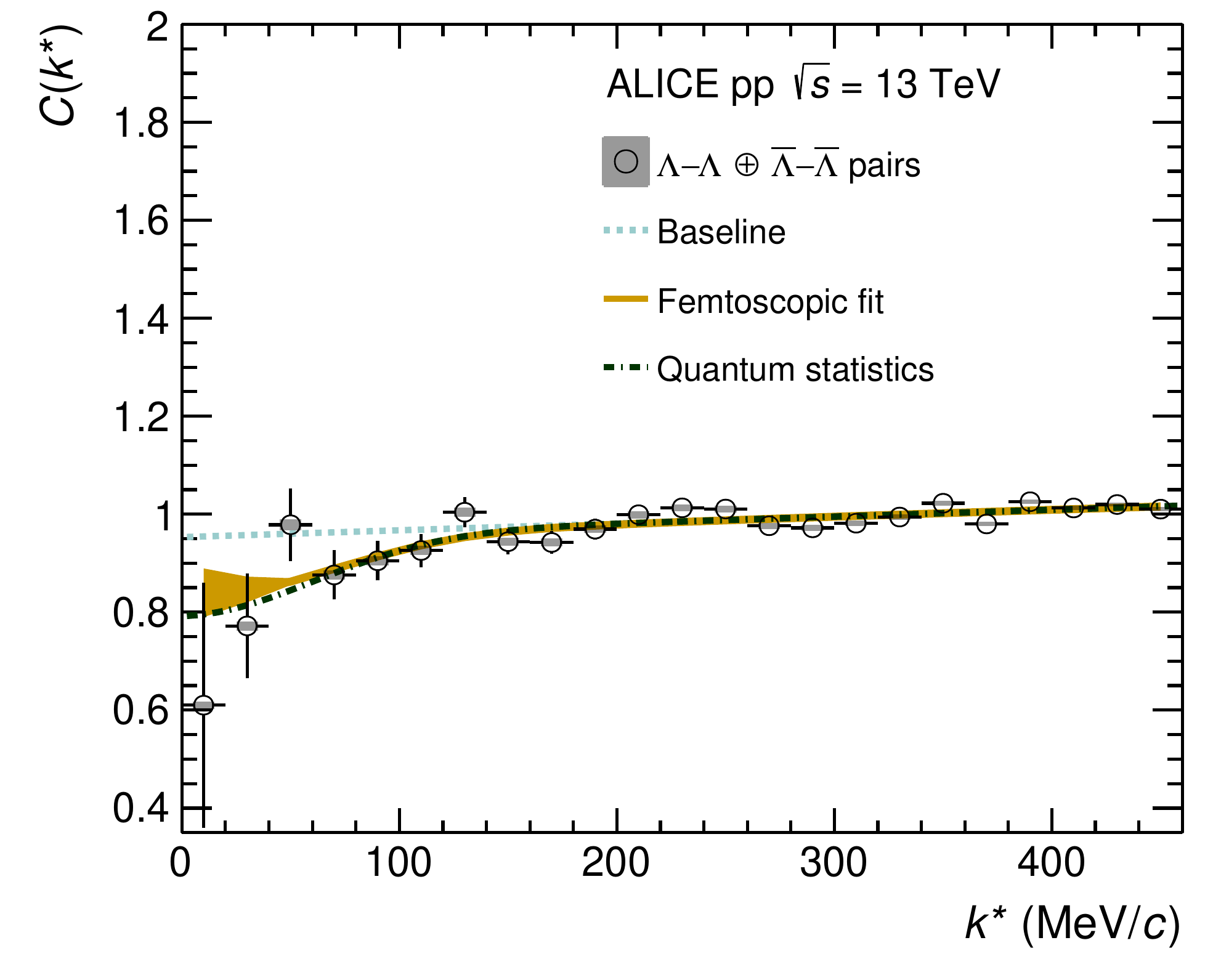}
}
\caption{Results for the fit of the {\ppColl} data at $\sqrt{s}=13$~TeV. The {\pp} correlation function (left panel) is fitted with CATS (blue line) and the {\LL} correlation function (right panel) is fitted with the Lednick\'y model (yellow line). The dashed line represents the linear baseline from Eq. \ref{eq:fit}, while the dark dashed-dotted line on top of the {\LL} data shows the expected correlation based on quantum statistics alone, in case of a strong interaction potential compatible with zero.
}
\label{fig:Ck_pp}
\end{figure*}

After all corrections have been applied to $C_\mathrm{tot}(k^*)$, the final fit function is obtained by multiplying it with a linear baseline $(a+b k^*)$ describing the normalization and non-femtoscopy background \cite{RUN1} 
\begin{equation}\label{eq:fit}
 C_\mathrm{fit}(k^*)=(a+b k^*) C_\mathrm{tot}(k^*).
\end{equation}
Figure~\ref{fig:Ck_pp} shows an example of the {\pp} and {\LL} correlation functions measured in {\ppColl} collisions at $\sqrt{s}=13$~TeV, together with the fit functions.
The {\pp} experimental data show a flat behaviour in the range $200<k^*<400$~MeV/$c$, thus by default the slope of the baseline is assumed to be zero ($b=0$) and the correlation is fitted in the range $k^*<375$~MeV/$c$. The resulting $r_0$ values are $1.182\pm0.008\text{(stat)}^{+0.005}_{-0.002}\text{(syst)}$~fm in {\ppColl} collisions at $\sqrt{s}=13$~TeV and $1.427\pm0.007\text{(stat)}^{+0.001}_{-0.014}\text{(syst)}$~fm in {\pPbColl} collisions at $\sqrt{s_\mathrm{NN}}=5.02$~TeV. In {\ppColl} collisions at $\sqrt{s}=7$~TeV the source size is $r_0=1.125\pm0.018\text{(stat)}^{+0.058}_{-0.035}\text{(syst)}$~fm~\cite{RUN1}.

The systematic uncertainties of the radius $r_0$ are evaluated following the prescription established during the analysis of {\ppColl} collisions at $\sqrt{s}=$7~TeV~\cite{RUN1}. The upper limit of the fit range for the {\pp} pairs is varied within $k^*\in\{350,375,400\}$~MeV/$c$ and the input to the $\lambda$ parameters is modified by 20\%, keeping primary and secondary fractions constant. 

Two further systematic variations are performed for the {\pp} correlation. The first concerns the possible effect of non-femtoscopy contributions to the correlation functions, which can be modeled by a linear baseline (see Eq.~\ref{eq:fit}) with the inclusion of $b$ as a free fit parameter. The final systematic variation is to model the {\pL} feed-down contribution by using a leading-order (LO) \cite{Polinder:2006zh,Haidenbauer:2013oca} computation to model the interaction. The effect of the latter is negligible, as the transformation to the {\pp} system smears the differences observed in the pure {\pL} correlation function out.

To investigate the {\LL} interaction the source sizes are fixed to the above results and the {\LL} correlations from all three data sets are fitted simultaneously in order to extract the scattering parameters. The correlation functions show a slight non-flat behaviour at large $k^*$, especially for the {\ppColl} collisions at $\sqrt{s}=13$~TeV (right panel in Fig.~\ref{fig:Ck_pp}). Thus the fit is performed by allowing a non-zero slope parameter $b$ (see Eq.~\ref{eq:fit}). The fit range is extended to $k^*<460$~MeV/$c$ in order to better constrain the linear baseline. Due to the small primary $\lambda$ parameters (see Table \ref{tab:lambdaval}) the {\LL} correlation signal is quite weak and the fit shows a slight systematic enhancement compared to the expected $C_\mathrm{tot}(k^*)$ due to quantum statistics only, suggestive of an attractive interaction. However, the current statistical uncertainties do not allow the {\LL} scattering parameters to be extracted from the fit. Therefore, an alternative approach to study the {\LL} interaction will be presented in the next section. Systematic uncertainties related to the {\LL} emission source may arise from several different effects, which are discussed in the rest of this section.

Previous studies have revealed that the emission source can be elongated along some of the spatial directions and have a multiplicity or $m_\mathrm{T}$ dependence \cite{Aamodt:2011kd,Sirunyan:2017ies}. In the present analysis it is assumed that the correlation function can be modeled by an effective Gaussian source. The validity of this statement is verified by a simple toy Monte Carlo, in which a data-driven multiplicity dependence is introduced into the source function and the resulting theoretical {\pp} correlation function computed with CATS. The deviations between this result and a correlation function obtained with an effective Gaussian source profile are negligible.

Possible differences in the effective emitting sources of {\pp} and {\LL} pairs due to the strong decays of broad resonances and $m_\mathrm{T}$ scaling are evaluated via simulations and estimated to have at most a $5$\% effect on the effective source size $r_0$. This is taken into account by including an additional systematic uncertainty on the $r_\text{\LL}$ value extracted from the fit to the {\pp} correlation. 
\section{Results}
\noindent
In order to extract the {\LL} scattering parameters, the correlation functions measured in {\ppColl} collisions at $\sqrt{s}=$7, 13~TeV as well as in {\pPbColl} collisions at $\sqrt{s_\mathrm{NN}}=5.02$~TeV are fitted simultaneously. The right panel in Fig.~{\ref{fig:Ck_pp}} shows the {\LL} correlation function obtained in {\ppColl} collisions at $\sqrt{s}=13$~TeV together with the result from the fit.  
\begin{figure*}[h]
\centering{
\includegraphics[width=0.49\textwidth]{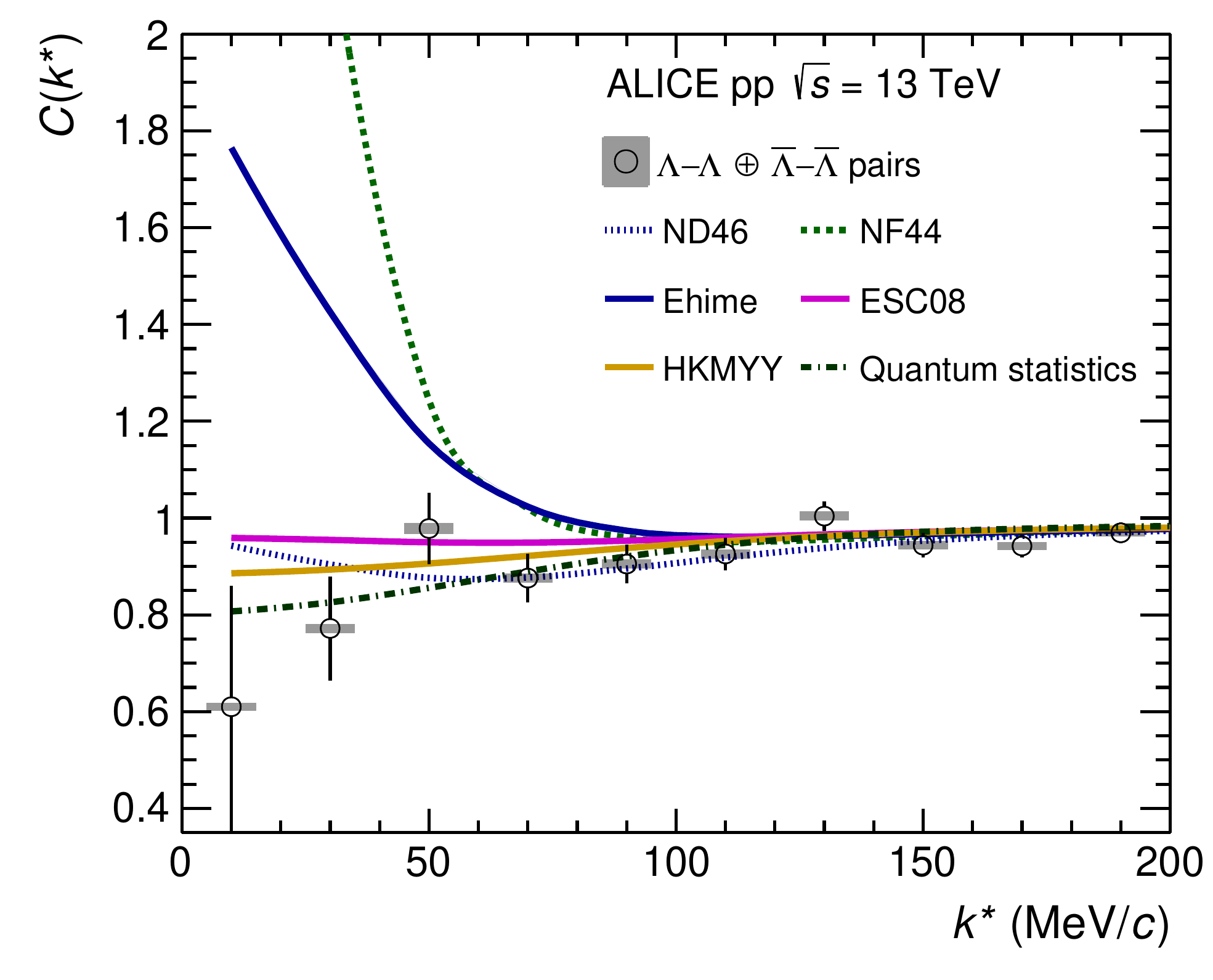}
\includegraphics[width=0.49\textwidth]{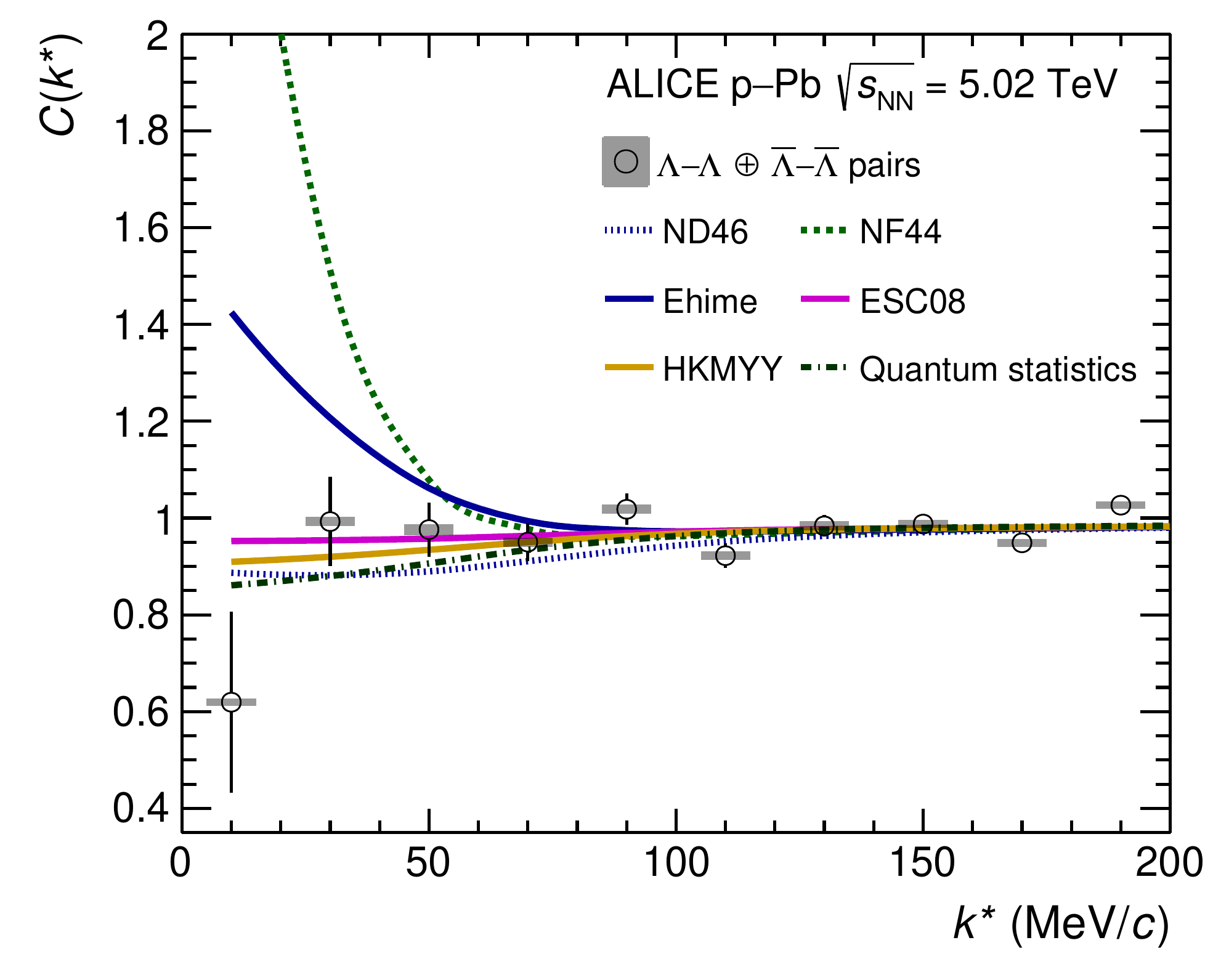}
}
\caption{{\LL} correlations measured in
{\ppColl} collisions at $\sqrt{s}=13$ TeV (left panel) and {\pPbColl} collisions at $\sqrt{s_\mathrm{NN}}=5.02$~TeV (right panel) together with the functions computed by the different models~\cite{Morita:2014kza}. The tested potentials are converted to correlation functions using CATS and the baseline is refitted for each model. The effects of momentum resolution and residuals are included in the theory curves.}
\label{fig:LL_mod_NEW}
\end{figure*}
\\
Since the uncertainties of the scattering parameters are large, different model predictions are tested on the basis of their agreement with the measured correlation functions.

One option is to use a local potential and obtain $C(k^*)$ based on the exact solution from CATS, with the source size fixed to the value obtained from the fit to the {\pp} correlations. Many of the existing model predictions are summarized in~\cite{Morita:2014kza} and the corresponding potentials $V(r)$ are parametrized in a local form using a double-Gaussian function. The correlation function depends on the nature of the underlying interaction and Fig.~\ref{fig:LL_mod_NEW} shows the experimental {\LL} correlations measured in {\ppColl} collisions at $\sqrt{s}=13$ TeV (left panel) and {\pPbColl} collisions at $\sqrt{s_\mathrm{NN}}=5.02$ TeV (right panel) together with the correlation functions obtained for different meson-exchange interaction potentials employing CATS. Models with a strongly attractive interaction ($f^{-1}_0\lesssim1$ and positive), like the Ehime \cite{Ueda:1998bz} potential, result in a large enhancement of the correlation function at low momenta which overshoots the data significantly both in {\ppColl} and {\pPbColl} collisions. The same is valid for potentials corresponding to a shallow bound state ($f^{-1}_0\rightarrow 0$ and negative), e.g. NF44 \cite{Nagels:1978sc}.

The other tested potentials correspond either to a bound state or a shallow attractive ($f^{-1}_0\gtrsim1$) non-binding interaction. However, those two very different scenarios result in similar correlations and are difficult to separate. This is evident from Fig.~\ref{fig:LL_mod_NEW} as all of the ESC08 \cite{Rijken:2010zzb}, HKMYY \cite{Hiyama:2002yj} and Nijmegen ND46 \cite{Nagels:1976xq} models produce comparable results and are compatible with the experimental data, even though their scattering parameters are different. In particular, ND46 predicts a bound state, while the ESC08 and HKMYY models describe a shallow attractive potential and the latter is consistent with hypernuclei data \cite{Takahashi:2001nm,Ahn:2013poa}.

The Lednick\'y model can be used to compute $C(k^*)$ for any $f^{-1}_0$ and $d_0$. Thus a scan over the scattering parameters can be preformed and the agreement to the experimental data can be quantified. The Lednick\'y model breaks down for source sizes smaller than the effective range, especially when dealing with repulsive interactions \cite{RUN1}, as it produces unphysical negative correlation functions. As there are no realistic models predicting such an interaction, this study is not affected. Nevertheless, all models described in \cite{Morita:2014kza} are explicitly tested by comparing the correlation functions obtained using the exact solution provided by CATS with the approximate solution evaluated using the Lednick\'y model. The deviations are on the percent level and are neglected.

\begin{figure*}[h]
\centering{
\includegraphics[width=0.75\textwidth]{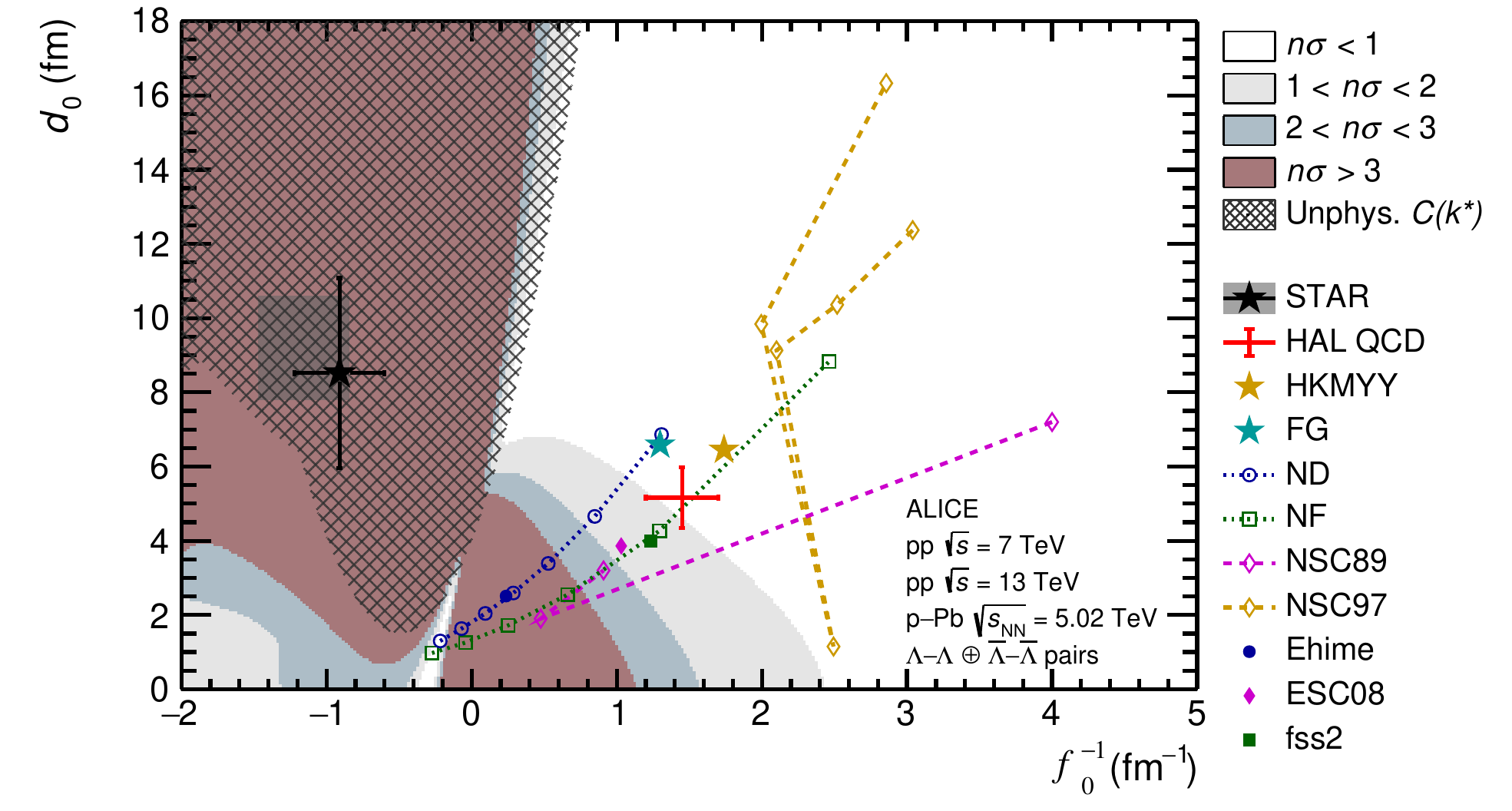}
}
 \caption{Exclusion plot for the {\LL} scattering parameters obtained using the {\LL} correlations from {\ppColl} collisions at $\sqrt{s}=7$ and 13~TeV as well as {\pPbColl} collisions at $\sqrt{s_{\mathrm{NN}}}=5.02$~TeV. The different colors represent the confidence level of excluding a set of parameters, given in $n\sigma$. The black hashed region is where the Lednick\'y model produces an unphysical correlation. The two models denoted by colored stars are compatible with hypernuclei data, while the red cross corresponds to the preliminary result of the lattice computation performed by the HAL QCD collaboration. For details regarding the region at slightly negative $f^{-1}_0$ and $d_0<4$, compatible with a bound state, refer to Fig.~\ref{fig:ExclusionEbin}.}
\label{fig:Exclusion}
\end{figure*}

Another assumption, which the Lednick\'y model is based on, is a Gaussian profile of the source. The EPOS \cite{Pierog:2013ria} transport model predicts a non-Gaussian emission profile \cite{CATS}, and the effects of short lived resonances are included. This source was adopted in CATS, by tuning its width such as to describe the {\pp} correlation function, and the predicted $C(k^*)$ for all of the ND and NF models, shown in Fig.~\ref{fig:Exclusion}, were compared to the {\LL} correlation function in {\ppColl} collisions at $\sqrt{s}=13$~TeV. The deviations in $\chi^2$ compared to the case of a Gaussian source are within the uncertainty, justifying the use of a Gaussian source.

To quantify the uncertainties of $f^{-1}_0$ and $d_0$, and estimate the confidence level of each parameter set, a Monte Carlo method is used. In the current work the approach described in \cite{NumericalRecipes} is followed, which is closely related to the Bootstrap method. The strategy is to use the Lednick\'y model to perform a scan over the parameter space spanned by $f^{-1}_0\in [-2,5]$~fm$^{-1}$ and $d_0\in [0,18]$~fm and refit the {\LL} correlation using Eq.~\ref{eq:fit} when fixing the scattering parameters to a specific value $(f^{-1}_0,d_0)_{i}$. The corresponding $\chi^2_{i}$ is evaluated by taking all data sets ({\ppColl} at $\sqrt{s}=7$ and $13$~TeV and {\pPbColl} at $\sqrt{s_{\mathrm{NN}}}=5.02$~TeV) into account. The different scattering parameters can be compared by finding the lowest (best) $\chi^2_\text{best}$ and evaluating $\Delta\chi^2_{i}=\chi^2_{i}-\chi^2_\text{best}$ for each parameter set. This observable, and the associated $(f^{-1}_0,d_0)_{i}$, can be directly linked to the confidence level~\cite{NumericalRecipes}. This can be achieved either by assuming normally distributed uncertainties of $(f^{-1}_0,d_0)$, or invoking a more sophisticated Monte Carlo study, like the Bootstrap method. The latter is used in the current analysis.

The resulting exclusion plot is presented in Fig.~\ref{fig:Exclusion}, where the color code corresponds to the confidence level $n\sigma$ for a specific choice of scattering parameters. In the computation only the statistical uncertainties are taken into account, as the systematic uncertainties are negligible according to the Barlow criterion \cite{Barlow}. The predicted scattering parameters of all discussed potentials are highlighted with different markers and the phase space region in which the Lednick\'y model produces an unphysical correlation is specified by the black hatched area. In this region the effective range expansion breaks down and the Lednick\'y equation leads to a negative correlation function. While the STAR result~\cite{Adamczyk:2014vca} is located in this region, all theoretical models exclude the possibility of a repulsive {\LL} interaction with large effective range. Moreover a re-analysis of the STAR data~\cite{Morita:2014kza} demonstrated that a more realistic treatment of the residual correlations leads to an inversion of the sign of the scattering length, that corresponds to an attractive potential. 
The imposed limit on the scattering length is $f_0^{-1}>0.8~$fm$^{-1}$~\cite{Morita:2014kza}. This result can be tested within the current work, and Fig.~\ref{fig:Exclusion} demonstrates that the ALICE data can extend those constraints. In particular the region corresponding to a strongly attractive or a very weakly binding short-range interaction (small $|f^{-1}_0|$ and small $d_0$) is excluded by the data, while a shallow attractive potential (large $f^{-1}_0$) is in very good agreement with the experimental results obtained from this analysis. A {\LL} bound state would correspond to negative $f^{-1}_0$ and small $d_0$ values. The present data are compatible with such a scenario, but the available phase space is strongly constrained. 
The HKMYY \cite{Hiyama:2002yj}, FG \cite{Filikhin:2002wm} and HAL QCD \cite{LatticePrivate} values are of particular interest, as the first two models are tuned to describe the modern hypernuclei data, while the latter is the latest state-of-the-art lattice computation from the HAL QCD collaboration. 
The lattice results are preliminary and predict the scattering parameters $f^{-1}_0=1.45\pm0.25$~fm$^{-1}$ and $d_0=5.16\pm0.82$~fm \cite{LatticePrivate}. All three models are compatible with the ALICE data, providing further support for a shallow attractive {\LL} interaction potential.
\begin{figure*}[h]
\centering{
\includegraphics[width=0.75\textwidth]{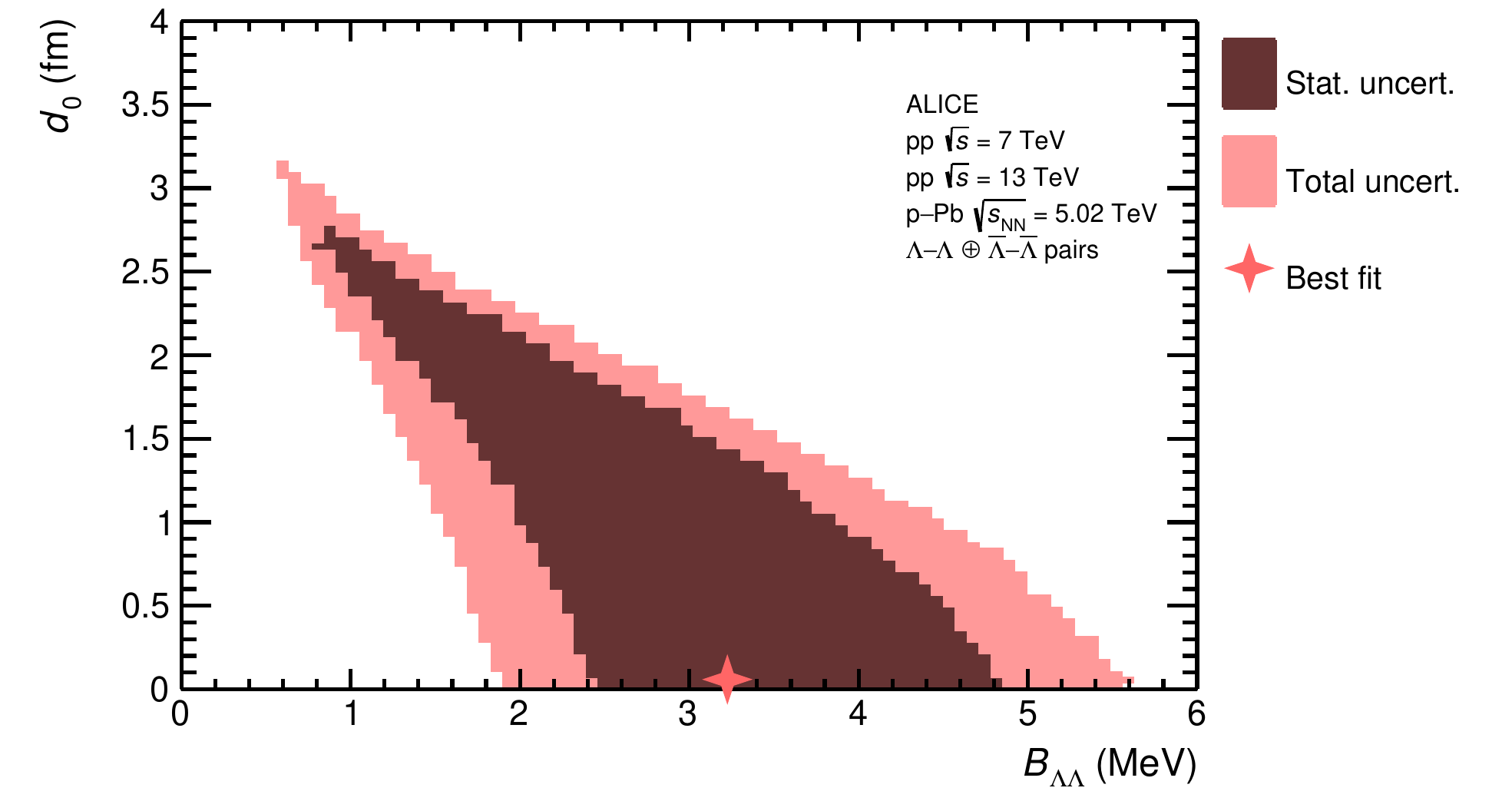}
}
\caption{The region of the 1$\sigma$ confidence level from Fig.~\ref{fig:Exclusion}, displayed in the ($B_{\Lambda\Lambda},d_0$) plane. The inner (dark) region corresponds to the statistical uncertainty of the method, while the outer (light) region includes the systematic variations. The red star corresponds to the parameters with the lowest $\chi^2$.}
\label{fig:ExclusionEbin}
\end{figure*}
\\
A possible bound state is investigated within the effective-range expansion by computing the corresponding binding energy from the relation \cite{Gongyo:2017fjb,Naidon:2016dpf}
\begin{equation}\label{eq:BE}
    B_{\Lambda \Lambda}=\dfrac{1}{m_\Lambda d_0 ^2}\left ( 1-\sqrt{1+2 d_0f^{-1}_0}\right)^2.
\end{equation}
This relation is only valid for bound states, which are characterized by negative $f^{-1}_0$ values. Further, the binding energy has to be a real number, thus the expression $1+2d_0f^{-1}_0$ has to be positive, which implies that at least one of the parameters $f^{-1}_0$ or $d_0$ has to be small in absolute value. With these restrictions Eq.~\ref{eq:BE} transforms the observables in the exclusion plot (Fig.~\ref{fig:Exclusion}) from $(f^{-1}_0,d_0)$ to $(B_{\Lambda\Lambda},d_0)$, considering only the parameter space compatible with a bound state. This is done in Fig.~\ref{fig:ExclusionEbin}, where only the 1$\sigma$ confidence region is shown, as it corresponds to the uncertainty of $B_{\Lambda\Lambda}$. The dark region marks the statistical uncertainty of the fit. The allowed binding energy, independent of $d_0$, is \EbinResultSTAT, where the central value corresponds to the lowest $\chi^2$ and the uncertainties are determined based on the lowest and highest allowed $B_{\Lambda\Lambda}$ values within the 1$\sigma$ confidence region. However the systematic uncertainties related to the source sizes are not taken into account, neither any possible biases related to the fit procedure. Thus the computation of the exclusion plots (Figs.~\ref{fig:Exclusion} and \ref{fig:ExclusionEbin}) was repeated 121 times, where in each re-iteration the source sizes related to the data sets are varied within the associated uncertainties, the fit ranges within $k^*\in\{420,460,500\}~$MeV/$c$ and the bin widths of the experimental correlations are chosen as 12, 16 and 20~MeV/$c$. The resulting fluctuations of the 1$\sigma$ confidence region are marked in Fig.~\ref{fig:ExclusionEbin} by the light region and represent the total uncertainty. Assuming the latter is the quadratic sum of the statistical and systematic uncertainty, the final result is \EbinResult.

\section{Summary}
\noindent
In this Letter, new data on {\pp} and {\LL} correlations in {\ppColl} collisions at $\sqrt{s}=13$~TeV and {\pPbColl} collisions at $\sqrt{s_\mathrm{NN}}=5.02$~TeV are presented. Together with the results from a pioneering study on two-baryon correlations in {\ppColl} at $\sqrt{s}=7$ TeV, these data allow for a detailed study of the {\LL} interaction with unprecedented precision.

Each data set is analyzed separately by extracting the {\pp} and {\LL} correlation functions. The former are used to constrain the size of the source $r_0$, which is assumed to be the same for {\pp} and {\LL} pairs. The {\LL} interaction is then investigated by testing the combined compatibility of all data sets to different model predictions and scattering parameters. The HKMYY and FG models, which are tuned to hypernuclei data, and the lattice calculations performed by the HAL QCD collaboration predict a shallow attractive interaction potential. The ALICE data manifest very good agreement with these predictions. Nevertheless, the data is also compatible with the existence of a bound state, given a binding energy of \EbinResult. The Run 3 of the LHC is expected to further increase the statistical significance of the {\LL} correlation function and allow the scattering parameters to be constraint even more precisely in the future.

\newenvironment{acknowledgement}{\relax}{\relax}
\begin{acknowledgement}
\section*{Acknowledgements}
The ALICE collaboration is grateful to the HAL QCD collaboration for providing lattice results regarding the {\LL} interaction. We are particularly thankful to Prof. Tetsuo Hatsuda and Prof. Kenji Sasaki for the precious suggestions and stimulating discussions.


The ALICE Collaboration would like to thank all its engineers and technicians for their invaluable contributions to the construction of the experiment and the CERN accelerator teams for the outstanding performance of the LHC complex.
The ALICE Collaboration gratefully acknowledges the resources and support provided by all Grid centres and the Worldwide LHC Computing Grid (WLCG) collaboration.
The ALICE Collaboration acknowledges the following funding agencies for their support in building and running the ALICE detector:
A. I. Alikhanyan National Science Laboratory (Yerevan Physics Institute) Foundation (ANSL), State Committee of Science and World Federation of Scientists (WFS), Armenia;
Austrian Academy of Sciences, Austrian Science Fund (FWF): [M 2467-N36] and Nationalstiftung f\"{u}r Forschung, Technologie und Entwicklung, Austria;
Ministry of Communications and High Technologies, National Nuclear Research Center, Azerbaijan;
Conselho Nacional de Desenvolvimento Cient\'{\i}fico e Tecnol\'{o}gico (CNPq), Universidade Federal do Rio Grande do Sul (UFRGS), Financiadora de Estudos e Projetos (Finep) and Funda\c{c}\~{a}o de Amparo \`{a} Pesquisa do Estado de S\~{a}o Paulo (FAPESP), Brazil;
Ministry of Science \& Technology of China (MSTC), National Natural Science Foundation of China (NSFC) and Ministry of Education of China (MOEC) , China;
Croatian Science Foundation and Ministry of Science and Education, Croatia;
Centro de Aplicaciones Tecnol\'{o}gicas y Desarrollo Nuclear (CEADEN), Cubaenerg\'{\i}a, Cuba;
Ministry of Education, Youth and Sports of the Czech Republic, Czech Republic;
The Danish Council for Independent Research | Natural Sciences, the Carlsberg Foundation and Danish National Research Foundation (DNRF), Denmark;
Helsinki Institute of Physics (HIP), Finland;
Commissariat \`{a} l'Energie Atomique (CEA), Institut National de Physique Nucl\'{e}aire et de Physique des Particules (IN2P3) and Centre National de la Recherche Scientifique (CNRS) and R\'{e}gion des  Pays de la Loire, France;
Bundesministerium f\"{u}r Bildung und Forschung (BMBF) and GSI Helmholtzzentrum f\"{u}r Schwerionenforschung GmbH, Germany;
General Secretariat for Research and Technology, Ministry of Education, Research and Religions, Greece;
National Research, Development and Innovation Office, Hungary;
Department of Atomic Energy Government of India (DAE), Department of Science and Technology, Government of India (DST), University Grants Commission, Government of India (UGC) and Council of Scientific and Industrial Research (CSIR), India;
Indonesian Institute of Science, Indonesia;
Centro Fermi - Museo Storico della Fisica e Centro Studi e Ricerche Enrico Fermi and Istituto Nazionale di Fisica Nucleare (INFN), Italy;
Institute for Innovative Science and Technology , Nagasaki Institute of Applied Science (IIST), Japan Society for the Promotion of Science (JSPS) KAKENHI and Japanese Ministry of Education, Culture, Sports, Science and Technology (MEXT), Japan;
Consejo Nacional de Ciencia (CONACYT) y Tecnolog\'{i}a, through Fondo de Cooperaci\'{o}n Internacional en Ciencia y Tecnolog\'{i}a (FONCICYT) and Direcci\'{o}n General de Asuntos del Personal Academico (DGAPA), Mexico;
Nederlandse Organisatie voor Wetenschappelijk Onderzoek (NWO), Netherlands;
The Research Council of Norway, Norway;
Commission on Science and Technology for Sustainable Development in the South (COMSATS), Pakistan;
Pontificia Universidad Cat\'{o}lica del Per\'{u}, Peru;
Ministry of Science and Higher Education and National Science Centre, Poland;
Korea Institute of Science and Technology Information and National Research Foundation of Korea (NRF), Republic of Korea;
Ministry of Education and Scientific Research, Institute of Atomic Physics and Ministry of Research and Innovation and Institute of Atomic Physics, Romania;
Joint Institute for Nuclear Research (JINR), Ministry of Education and Science of the Russian Federation, National Research Centre Kurchatov Institute, Russian Science Foundation and Russian Foundation for Basic Research, Russia;
Ministry of Education, Science, Research and Sport of the Slovak Republic, Slovakia;
National Research Foundation of South Africa, South Africa;
Swedish Research Council (VR) and Knut \& Alice Wallenberg Foundation (KAW), Sweden;
European Organization for Nuclear Research, Switzerland;
National Science and Technology Development Agency (NSDTA), Suranaree University of Technology (SUT) and Office of the Higher Education Commission under NRU project of Thailand, Thailand;
Turkish Atomic Energy Agency (TAEK), Turkey;
National Academy of  Sciences of Ukraine, Ukraine;
Science and Technology Facilities Council (STFC), United Kingdom;
National Science Foundation of the United States of America (NSF) and United States Department of Energy, Office of Nuclear Physics (DOE NP), United States of America.    
\end{acknowledgement}

\bibliographystyle{utphys}   
\bibliography{LambdaLambdaPaper}

\newpage
\appendix
\section{The ALICE Collaboration}
\label{app:collab}

\begingroup
\small
\begin{flushleft}
S.~Acharya\Irefn{org141}\And 
D.~Adamov\'{a}\Irefn{org93}\And 
S.P.~Adhya\Irefn{org141}\And 
A.~Adler\Irefn{org74}\And 
J.~Adolfsson\Irefn{org80}\And 
M.M.~Aggarwal\Irefn{org98}\And 
G.~Aglieri Rinella\Irefn{org34}\And 
M.~Agnello\Irefn{org31}\And 
N.~Agrawal\Irefn{org10}\And 
Z.~Ahammed\Irefn{org141}
% ...
\And
\renewcommand\labelenumi{\textsuperscript{\theenumi}~}

\section*{Affiliation notes}
\renewcommand\theenumi{\roman{enumi}}
\begin{Authlist}
\item \Adef{org*}Deceased
\item \Adef{orgI}Dipartimento DET del Politecnico di Torino, Turin, Italy
\item \Adef{orgII}M.V. Lomonosov Moscow State University, D.V. Skobeltsyn Institute of Nuclear, Physics, Moscow, Russia
\item \Adef{orgIII}Department of Applied Physics, Aligarh Muslim University, Aligarh, India
\item \Adef{orgIV}Institute of Theoretical Physics, University of Wroclaw, Poland
\end{Authlist}

\section*{Collaboration Institutes}
\renewcommand\theenumi{\arabic{enumi}~}
\begin{Authlist}
\item \Idef{org1}A.I. Alikhanyan National Science Laboratory (Yerevan Physics Institute) Foundation, Yerevan, Armenia
\item \Idef{org2}Bogolyubov Institute for Theoretical Physics, National Academy of Sciences of Ukraine, Kiev, Ukraine
\item \Idef{org3}Bose Institute, Department of Physics  and Centre for Astroparticle Physics and Space Science (CAPSS), Kolkata, India
\item \Idef{org4}Budker Institute for Nuclear Physics, Novosibirsk, Russia
\item \Idef{org5}California Polytechnic State University, San Luis Obispo, California, United States
\item \Idef{org6}Central China Normal University, Wuhan, China
\item \Idef{org7}Centre de Calcul de l'IN2P3, Villeurbanne, Lyon, France
\item \Idef{org8}Centro de Aplicaciones Tecnol\'{o}gicas y Desarrollo Nuclear (CEADEN), Havana, Cuba
\item \Idef{org9}Centro de Investigaci\'{o}n y de Estudios Avanzados (CINVESTAV), Mexico City and M\'{e}rida, Mexico
\item \Idef{org10}Centro Fermi - Museo Storico della Fisica e Centro Studi e Ricerche ``Enrico Fermi', Rome, Italy
\item \Idef{org11}Chicago State University, Chicago, Illinois, United States
\item \Idef{org12}China Institute of Atomic Energy, Beijing, China
\item \Idef{org13}Chonbuk National University, Jeonju, Republic of Korea
\item \Idef{org14}Comenius University Bratislava, Faculty of Mathematics, Physics and Informatics, Bratislava, Slovakia
\item \Idef{org15}COMSATS University Islamabad, Islamabad, Pakistan
\item \Idef{org16}Creighton University, Omaha, Nebraska, United States
\item \Idef{org17}Department of Physics, Aligarh Muslim University, Aligarh, India
\item \Idef{org18}Department of Physics, Pusan National University, Pusan, Republic of Korea
\item \Idef{org19}Department of Physics, Sejong University, Seoul, Republic of Korea
\item \Idef{org20}Department of Physics, University of California, Berkeley, California, United States
\item \Idef{org21}Department of Physics, University of Oslo, Oslo, Norway
\item \Idef{org22}Department of Physics and Technology, University of Bergen, Bergen, Norway
\item \Idef{org23}Dipartimento di Fisica dell'Universit\`{a} 'La Sapienza' and Sezione INFN, Rome, Italy
\item \Idef{org24}Dipartimento di Fisica dell'Universit\`{a} and Sezione INFN, Cagliari, Italy
\item \Idef{org25}Dipartimento di Fisica dell'Universit\`{a} and Sezione INFN, Trieste, Italy
\item \Idef{org26}Dipartimento di Fisica dell'Universit\`{a} and Sezione INFN, Turin, Italy
\item \Idef{org27}Dipartimento di Fisica e Astronomia dell'Universit\`{a} and Sezione INFN, Bologna, Italy
\item \Idef{org28}Dipartimento di Fisica e Astronomia dell'Universit\`{a} and Sezione INFN, Catania, Italy
\item \Idef{org29}Dipartimento di Fisica e Astronomia dell'Universit\`{a} and Sezione INFN, Padova, Italy
\item \Idef{org30}Dipartimento di Fisica `E.R.~Caianiello' dell'Universit\`{a} and Gruppo Collegato INFN, Salerno, Italy
\item \Idef{org31}Dipartimento DISAT del Politecnico and Sezione INFN, Turin, Italy
\item \Idef{org32}Dipartimento di Scienze e Innovazione Tecnologica dell'Universit\`{a} del Piemonte Orientale and INFN Sezione di Torino, Alessandria, Italy
\item \Idef{org33}Dipartimento Interateneo di Fisica `M.~Merlin' and Sezione INFN, Bari, Italy
\item \Idef{org34}European Organization for Nuclear Research (CERN), Geneva, Switzerland
\item \Idef{org35}Faculty of Electrical Engineering, Mechanical Engineering and Naval Architecture, University of Split, Split, Croatia
\item \Idef{org36}Faculty of Engineering and Science, Western Norway University of Applied Sciences, Bergen, Norway
\item \Idef{org37}Faculty of Nuclear Sciences and Physical Engineering, Czech Technical University in Prague, Prague, Czech Republic
\item \Idef{org38}Faculty of Science, P.J.~\v{S}af\'{a}rik University, Ko\v{s}ice, Slovakia
\item \Idef{org39}Frankfurt Institute for Advanced Studies, Johann Wolfgang Goethe-Universit\"{a}t Frankfurt, Frankfurt, Germany
\item \Idef{org40}Gangneung-Wonju National University, Gangneung, Republic of Korea
\item \Idef{org41}Gauhati University, Department of Physics, Guwahati, India
\item \Idef{org42}Helmholtz-Institut f\"{u}r Strahlen- und Kernphysik, Rheinische Friedrich-Wilhelms-Universit\"{a}t Bonn, Bonn, Germany
\item \Idef{org43}Helsinki Institute of Physics (HIP), Helsinki, Finland
\item \Idef{org44}High Energy Physics Group,  Universidad Aut\'{o}noma de Puebla, Puebla, Mexico
\item \Idef{org45}Hiroshima University, Hiroshima, Japan
\item \Idef{org46}Hochschule Worms, Zentrum  f\"{u}r Technologietransfer und Telekommunikation (ZTT), Worms, Germany
\item \Idef{org47}Horia Hulubei National Institute of Physics and Nuclear Engineering, Bucharest, Romania
\item \Idef{org48}Indian Institute of Technology Bombay (IIT), Mumbai, India
\item \Idef{org49}Indian Institute of Technology Indore, Indore, India
\item \Idef{org50}Indonesian Institute of Sciences, Jakarta, Indonesia
\item \Idef{org51}INFN, Laboratori Nazionali di Frascati, Frascati, Italy
\item \Idef{org52}INFN, Sezione di Bari, Bari, Italy
\item \Idef{org53}INFN, Sezione di Bologna, Bologna, Italy
\item \Idef{org54}INFN, Sezione di Cagliari, Cagliari, Italy
\item \Idef{org55}INFN, Sezione di Catania, Catania, Italy
\item \Idef{org56}INFN, Sezione di Padova, Padova, Italy
\item \Idef{org57}INFN, Sezione di Roma, Rome, Italy
\item \Idef{org58}INFN, Sezione di Torino, Turin, Italy
\item \Idef{org59}INFN, Sezione di Trieste, Trieste, Italy
\item \Idef{org60}Inha University, Incheon, Republic of Korea
\item \Idef{org61}Institut de Physique Nucl\'{e}aire d'Orsay (IPNO), Institut National de Physique Nucl\'{e}aire et de Physique des Particules (IN2P3/CNRS), Universit\'{e} de Paris-Sud, Universit\'{e} Paris-Saclay, Orsay, France
\item \Idef{org62}Institute for Nuclear Research, Academy of Sciences, Moscow, Russia
\item \Idef{org63}Institute for Subatomic Physics, Utrecht University/Nikhef, Utrecht, Netherlands
\item \Idef{org64}Institute for Theoretical and Experimental Physics, Moscow, Russia
\item \Idef{org65}Institute of Experimental Physics, Slovak Academy of Sciences, Ko\v{s}ice, Slovakia
\item \Idef{org66}Institute of Physics, Homi Bhabha National Institute, Bhubaneswar, India
\item \Idef{org67}Institute of Physics of the Czech Academy of Sciences, Prague, Czech Republic
\item \Idef{org68}Institute of Space Science (ISS), Bucharest, Romania
\item \Idef{org69}Institut f\"{u}r Kernphysik, Johann Wolfgang Goethe-Universit\"{a}t Frankfurt, Frankfurt, Germany
\item \Idef{org70}Instituto de Ciencias Nucleares, Universidad Nacional Aut\'{o}noma de M\'{e}xico, Mexico City, Mexico
\item \Idef{org71}Instituto de F\'{i}sica, Universidade Federal do Rio Grande do Sul (UFRGS), Porto Alegre, Brazil
\item \Idef{org72}Instituto de F\'{\i}sica, Universidad Nacional Aut\'{o}noma de M\'{e}xico, Mexico City, Mexico
\item \Idef{org73}iThemba LABS, National Research Foundation, Somerset West, South Africa
\item \Idef{org74}Johann-Wolfgang-Goethe Universit\"{a}t Frankfurt Institut f\"{u}r Informatik, Fachbereich Informatik und Mathematik, Frankfurt, Germany
\item \Idef{org75}Joint Institute for Nuclear Research (JINR), Dubna, Russia
\item \Idef{org76}Korea Institute of Science and Technology Information, Daejeon, Republic of Korea
\item \Idef{org77}KTO Karatay University, Konya, Turkey
\item \Idef{org78}Laboratoire de Physique Subatomique et de Cosmologie, Universit\'{e} Grenoble-Alpes, CNRS-IN2P3, Grenoble, France
\item \Idef{org79}Lawrence Berkeley National Laboratory, Berkeley, California, United States
\item \Idef{org80}Lund University Department of Physics, Division of Particle Physics, Lund, Sweden
\item \Idef{org81}Nagasaki Institute of Applied Science, Nagasaki, Japan
\item \Idef{org82}Nara Women{'}s University (NWU), Nara, Japan
\item \Idef{org83}National and Kapodistrian University of Athens, School of Science, Department of Physics , Athens, Greece
\item \Idef{org84}National Centre for Nuclear Research, Warsaw, Poland
\item \Idef{org85}National Institute of Science Education and Research, Homi Bhabha National Institute, Jatni, India
\item \Idef{org86}National Nuclear Research Center, Baku, Azerbaijan
\item \Idef{org87}National Research Centre Kurchatov Institute, Moscow, Russia
\item \Idef{org88}Niels Bohr Institute, University of Copenhagen, Copenhagen, Denmark
\item \Idef{org89}Nikhef, National institute for subatomic physics, Amsterdam, Netherlands
\item \Idef{org90}NRC Kurchatov Institute IHEP, Protvino, Russia
\item \Idef{org91}NRNU Moscow Engineering Physics Institute, Moscow, Russia
\item \Idef{org92}Nuclear Physics Group, STFC Daresbury Laboratory, Daresbury, United Kingdom
\item \Idef{org93}Nuclear Physics Institute of the Czech Academy of Sciences, \v{R}e\v{z} u Prahy, Czech Republic
\item \Idef{org94}Oak Ridge National Laboratory, Oak Ridge, Tennessee, United States
\item \Idef{org95}Ohio State University, Columbus, Ohio, United States
\item \Idef{org96}Petersburg Nuclear Physics Institute, Gatchina, Russia
\item \Idef{org97}Physics department, Faculty of science, University of Zagreb, Zagreb, Croatia
\item \Idef{org98}Physics Department, Panjab University, Chandigarh, India
\item \Idef{org99}Physics Department, University of Jammu, Jammu, India
\item \Idef{org100}Physics Department, University of Rajasthan, Jaipur, India
\item \Idef{org101}Physikalisches Institut, Eberhard-Karls-Universit\"{a}t T\"{u}bingen, T\"{u}bingen, Germany
\item \Idef{org102}Physikalisches Institut, Ruprecht-Karls-Universit\"{a}t Heidelberg, Heidelberg, Germany
\item \Idef{org103}Physik Department, Technische Universit\"{a}t M\"{u}nchen, Munich, Germany
\item \Idef{org104}Politecnico di Bari, Bari, Italy
\item \Idef{org105}Research Division and ExtreMe Matter Institute EMMI, GSI Helmholtzzentrum f\"ur Schwerionenforschung GmbH, Darmstadt, Germany
\item \Idef{org106}Rudjer Bo\v{s}kovi\'{c} Institute, Zagreb, Croatia
\item \Idef{org107}Russian Federal Nuclear Center (VNIIEF), Sarov, Russia
\item \Idef{org108}Saha Institute of Nuclear Physics, Homi Bhabha National Institute, Kolkata, India
\item \Idef{org109}School of Physics and Astronomy, University of Birmingham, Birmingham, United Kingdom
\item \Idef{org110}Secci\'{o}n F\'{\i}sica, Departamento de Ciencias, Pontificia Universidad Cat\'{o}lica del Per\'{u}, Lima, Peru
\item \Idef{org111}Shanghai Institute of Applied Physics, Shanghai, China
\item \Idef{org112}St. Petersburg State University, St. Petersburg, Russia
\item \Idef{org113}Stefan Meyer Institut f\"{u}r Subatomare Physik (SMI), Vienna, Austria
\item \Idef{org114}SUBATECH, IMT Atlantique, Universit\'{e} de Nantes, CNRS-IN2P3, Nantes, France
\item \Idef{org115}Suranaree University of Technology, Nakhon Ratchasima, Thailand
\item \Idef{org116}Technical University of Ko\v{s}ice, Ko\v{s}ice, Slovakia
\item \Idef{org117}Technische Universit\"{a}t M\"{u}nchen, Excellence Cluster 'Universe', Munich, Germany
\item \Idef{org118}The Henryk Niewodniczanski Institute of Nuclear Physics, Polish Academy of Sciences, Cracow, Poland
\item \Idef{org119}The University of Texas at Austin, Austin, Texas, United States
\item \Idef{org120}Universidad Aut\'{o}noma de Sinaloa, Culiac\'{a}n, Mexico
\item \Idef{org121}Universidade de S\~{a}o Paulo (USP), S\~{a}o Paulo, Brazil
\item \Idef{org122}Universidade Estadual de Campinas (UNICAMP), Campinas, Brazil
\item \Idef{org123}Universidade Federal do ABC, Santo Andre, Brazil
\item \Idef{org124}University College of Southeast Norway, Tonsberg, Norway
\item \Idef{org125}University of Cape Town, Cape Town, South Africa
\item \Idef{org126}University of Houston, Houston, Texas, United States
\item \Idef{org127}University of Jyv\"{a}skyl\"{a}, Jyv\"{a}skyl\"{a}, Finland
\item \Idef{org128}University of Liverpool, Liverpool, United Kingdom
\item \Idef{org129}University of Science and Techonology of China, Hefei, China
\item \Idef{org130}University of Tennessee, Knoxville, Tennessee, United States
\item \Idef{org131}University of the Witwatersrand, Johannesburg, South Africa
\item \Idef{org132}University of Tokyo, Tokyo, Japan
\item \Idef{org133}University of Tsukuba, Tsukuba, Japan
\item \Idef{org134}Universit\'{e} Clermont Auvergne, CNRS/IN2P3, LPC, Clermont-Ferrand, France
\item \Idef{org135}Universit\'{e} de Lyon, Universit\'{e} Lyon 1, CNRS/IN2P3, IPN-Lyon, Villeurbanne, Lyon, France
\item \Idef{org136}Universit\'{e} de Strasbourg, CNRS, IPHC UMR 7178, F-67000 Strasbourg, France, Strasbourg, France
\item \Idef{org137}Universit\'{e} Paris-Saclay Centre d'Etudes de Saclay (CEA), IRFU, D\'{e}partment de Physique Nucl\'{e}aire (DPhN), Saclay, France
\item \Idef{org138}Universit\`{a} degli Studi di Foggia, Foggia, Italy
\item \Idef{org139}Universit\`{a} degli Studi di Pavia, Pavia, Italy
\item \Idef{org140}Universit\`{a} di Brescia, Brescia, Italy
\item \Idef{org141}Variable Energy Cyclotron Centre, Homi Bhabha National Institute, Kolkata, India
\item \Idef{org142}Warsaw University of Technology, Warsaw, Poland
\item \Idef{org143}Wayne State University, Detroit, Michigan, United States
\item \Idef{org144}Westf\"{a}lische Wilhelms-Universit\"{a}t M\"{u}nster, Institut f\"{u}r Kernphysik, M\"{u}nster, Germany
\item \Idef{org145}Wigner Research Centre for Physics, Hungarian Academy of Sciences, Budapest, Hungary
\item \Idef{org146}Yale University, New Haven, Connecticut, United States
\item \Idef{org147}Yonsei University, Seoul, Republic of Korea
\end{Authlist}
\endgroup
\end{document}